\documentclass[10pt,a4paper]{article}
\usepackage{graphicx}
\usepackage{tikz}
\usetikzlibrary{decorations.markings}
\usepackage{caption}
\usepackage{a4wide,epic}
\usepackage{subcaption}

\usepackage{amsmath,amssymb}
\usepackage{subcaption}
\usepackage[utf8]{inputenc}
\usepackage[T1]{fontenc}
\usepackage[colorlinks=true]{hyperref}

\newcommand{\tr}[1]{\operatorname{Tr}\left(#1\right)}

\newcommand{\bra}[1]{\langle #1 |}
\newcommand{\ket}[1]{| #1 \rangle}
\newcommand{\bket}[1]{\left\langle #1 \right\rangle}

\newcommand{\hrho}{\widehat{\rho}}

\newcommand{\RR}{{\mathbb R}}
\newcommand{\CC}{{\mathbb C}}

\newcommand{\bI}{\boldsymbol{I}}

\newcommand{\bP}{\boldsymbol{p}}

\newcommand{\minou}{\text{-}}

\newcommand{\cD}{\mathcal{D}}
\newcommand{\cM}{\mathcal{M}}

\newcommand{\Exp}{\mathbb{E}}

\newtheorem{thm}{Theorem}[section]
\newtheorem{lem}{Lemma}[section]

\newcommand{\pmax}{p_{\text{\tiny max}}}
\newcommand{\pmin}{p_{\text{\tiny min}}}

\begin{document}

\title{Exponential stabilization of quantum systems under continuous non-demolition measurements
\footnote{This work has been supported by the ANR project HAMROQS\newline 
$^{1}$Centre Automatique et Syst\`emes, Mines-ParisTech, PSL Research University. 60 Bd Saint-Michel, 75006 Paris, France.\newline
$^{2}$QUANTIC lab, INRIA Paris, rue Simone Iff 2, 75012 Paris, France.\newline
$^{3}$Department of Electronics and Information Systems, Ghent University, Belgium.\newline
{\tt\small gerardo.cardona@mines-paristech.fr, pierre.rouchon@mines-paristech.fr, (corresponding author) alain.sarlette@inria.fr}}
}

\author{Gerardo Cardona$^{1,2}$, Alain Sarlette$^{2,3}$ and Pierre Rouchon$^{1,2}$}  

\maketitle
		
%
%
%
		
\begin{abstract} 
We present a novel continuous-time control strategy to exponentially stabilize an eigenstate of a Quantum Non-Demolition (QND) measurement operator. In open-loop, the system converges to a random eigenstate of the measurement operator. The role of the feedback is to prepare a prescribed QND eigenstate with unit probability. To achieve this we introduce the use of Brownian motion to drive the unitary control actions; the feedback loop just adapts the amplitude of this Brownian noise input as a function of the system state. Essentially, it ``shakes'' the system away from undesired eigenstates by applying strong noise there, while relying on the open-loop dynamics to progressively reach the target. We prove exponential convergence towards the target eigenstate using standard stochastic Lyapunov methods.
The feedback scheme and its stability analysis suggest the use of an approximate filter which only tracks the populations of the eigenstates of the measurement operator. Such reduced filters should play an increasing role towards advanced quantum technologies.
\end{abstract}
		
%

\section{Introduction}


The progress of methods for measuring and controlling quantum systems  \cite{sayrin2011real,campagne2016using} now allows the physics community to implement building blocks of quantum information processors \cite{PhysRevLett.115.137002,reed2012realization,ofekQEC} that pose challenging control problems. One of the elementary building blocks is the stabilization of a quantum system onto a target eigenstate of a measurement operator. In a quantum computer \cite{nielsen2002quantum}, such stabilization could be used e.g.~for initializing the input states, or for providing auxiliary states that enable particular operations, like entangled states for quantum teleportation \cite{PhysRevLett.76.722,yamamoto2007feedback,PhysRevA.95.032329} and metrology \cite{PhysRevLett.116.093602} or magic states for performing T-gates \cite{PhysRevA.71.022316}. Moreover, quantum error correction protects information by encoding it in a larger state space and rejecting deviations from the nominal code-space \cite{gottesman1997stabilizer}. In this sense, strategies for stabilizing a state are a stepping stone towards stabilizing a code \emph{subspace} and thus protecting information towards quantum information processing. An extension of the present setting towards quantum error correction can be found in \cite{ahn2002continuous,cardona2019QEC}.

The defining element of our control setting is \emph{continuous} quantum measurement, where a continuous-time signal provides weak information, associated to weak backaction, on the quantum state \cite{barchielli1Belavkin991measurements}. In general, this setting allows to weakly measure non-commuting observables in parallel and to have measurement channels associated to non-Hermitian operators like the one describing energy loss \cite{PhysRevX.6.011002}. The most standard measurement for feedback control though is a Quantum non-demolition (QND) measurement \cite{HarocheBook}. As a close continuous-time counterpart to the projective measurement of a quantum observable, the measurement is characterized by a Hermitian operator; following stochastic dynamics, the system progressively converges towards one of the eigenstates of this measurement operator, in agreement with the stochastic measurement results. All eigenstates of the measurement operator are thus invariant under the dynamics, and as such the QND measurement itself can be considered as preparation tool \cite{dotsenko2009quantum}. Since the system converges to one of the eigenstates at random, an additional feedback control is necessary for preparing a particular target QND eigenstate. We here address how to add this feedback layer in continuous time.

Stabilization of QND eigenstates with continuous-time measurements has been investigated in the literature extensively \cite{stockton2004deterministic,van2005feedback,mirrahimi2007stabilizing,tsumura2008global,HadisCDC2018}. Our proposal is meant to improve these results on three aspects. First and foremost, while previous proposals prove convergence in a proper probabilistic sense, their analysis technique does not provide an estimate of the rate of convergence. By using a Lyapunov technique and a novel control approach, we here prove exponential convergence. This is not unexpected since the the QND measurement process in open loop converges exponentially towards the \textit{set} of QND eigenstates. Second, the feedback laws in existing work a priori depend on the full state $\rho$, which must be estimated in real-time. Our feedback law only depends on eigenstate populations which leads to a reduced filter. Third, existing feedback laws rely on conditional pulses, inspired by optimal discrete-time strategies, yet with the danger that such abrupt signals excite spurious dynamics in the fragile quantum system. Our strategy instead uses continuous control signals, consisting of Brownian noise whose gain is adapted as a function of the estimated state. We call this \textit{noise-assisted} quantum feedback.

In a nutshell, the strategy that we propose is rather simple to summarize. We let the system evolve in open loop until it gets close to one of the QND eigenstates. If it is the good one, we are done. If it is another one, as we get closer to it, we increase noise input on the system, effectively shaking it away from the bad eigenstate such that it has another chance to converge to the good one. This strategy works with generic conditions on the control Hamiltonian and simple control logic. The detail of the convergence proof is somewhat challenging, but transparent to the end user. The use of a noisy input signal can also be understood as a necessity to induce global exponential convergence on a compact set.

The structure of this paper is as follows: Section \ref{section:ProblemSetting} presents the dynamical model for QND measurements and introduces the control problem. Section \ref{section:PreviousWork} provides a brief overview of current feedback designs. In section \ref{section:NoiseAssistedFeedback} we introduce the use of Brownian motion to drive the controls and the standard control design that will be followed. Section \ref{section:ExpoStab} presents the main convergence theorem, proving exponential convergence towards a target eigenstate with a closed-loop Lyapunov function. In section \ref{section:Qfilters} we present an approximated filter to estimate the eigenstate populations, inspired from the stability analysis of the previous section. Lastly, in section \ref{section:Simulations} we make numerical simulations on a spin $J$ system, illustrating the robustness of the control approach.

\section{Open-loop QND dynamics and feedback goal}\label{section:ProblemSetting}

Consider a quantum system of finite dimension $n$. The state space is the set of density matrices $\mathcal{S}=\{\rho\in\CC^{n\times n}: \rho=\rho^\dagger,\rho \text{ positive semidefinite},\tr{\rho}=1 \}$. Here $\tr{\ \cdot \ }$ denotes the trace, $A^\dagger$ is the complex conjugate transpose of $A$.
The \textit{open-loop system} for a continuous-time quantum measurement with a single measurement channel is governed by the It\={o} stochastic differential equations~\cite{barchielli2009quantum}:
\begin{align}
\label{eq:QND_OL}d\rho_t&=\cD_{L}(\rho_t)dt+\sqrt{\eta}\cM_{L}(\rho_t)dW_t\; ,\\
\label{eq:QND_Y} dY_t &=\sqrt\eta\tr{(L+L^\dagger)\rho_t}dt+dW_t \;.
\end{align}
with the super-operators
\begin{align*}
    \cD_L(\rho)&=\big (L\rho L^\dagger-\tfrac{1}{2}(L^\dagger L \rho + \rho L^\dagger L)\big)
    \\
    \cM_L(\rho)&=\big(L\rho+\rho L^\dagger -\tr{\rho(L+L^\dagger)}\rho\big).
\end{align*}
Here $L\in\CC^{n\times n}$ is the measurement operator, $W$ is a standard Brownian motion, $Y\in\RR$ corresponds to the measurement process and $\eta \in [0,1]$ to its efficiency. The second equation describes the stochastic measurement output signal, the first one describes the corresponding measurement backaction. 

A QND measurement corresponds to a Hermitian measurement operator $L=L^\dag$. Consider the spectral decomposition $L=\sum_{k=1}^d \lambda_k\Pi_k$ where $\lambda_{1},...\lambda_d$ are the distinct ($d\leq n$), real eigenvalues of $L$ with corresponding orthogonal projection operators $\Pi_1,...,\Pi_d$ resolving the identity, i.e.~$\sum_{k=1}^d \Pi_k = \bI$. The population of the eigenspace $k$ is denoted by
\begin{equation}
 \label{eq:population}
 \bP_{k}(\rho_t):=\tr{\rho_t\Pi_k}\ge 0 \; ,
\end{equation}
with the property $\sum_{k=1}^d \;  \bP_{k}(\rho_t) =1$. 
The following Lemma summarizes the asymptotic behavior of~\eqref{eq:QND_OL} (without mentioning the corresponding measurement signal). It is based on an original exponential Lyapunov function $V_o$ providing an estimate of the  convergence rate towards the set of stationary states.
\begin{lem}\label{lemma:OpenLoop}
Consider the open-loop system \eqref{eq:QND_OL}with initial condition $\rho_0\in \mathcal{S}$. Then any realization of~\eqref{eq:QND_OL} remains in $\mathcal{S}$. Moreover	
\begin{enumerate}
  \item[(i)] For any $k$, the subspace population $\bP_{k}(\rho_t)$ is a martingale,  i.e. $\Exp[\bP_{k}(\rho_t) ]=\bP_{k}(\rho_0)$.
  \item[(ii)]  If there exists $k \in \{1,2,...,d\}$ such that $\bP_{k}(\rho_0)=1$, then $\rho_0$ is a steady state of \eqref{eq:QND_OL}.
   \item[(iii)]  The Lyapunov function
				$$ V_o(\rho) =\sum_{1\leq k< k' \leq d}\sqrt{\bP_k(\rho)}\sqrt{\bP_{k'}(\rho)},$$ decreases exponentially as
				$$\quad\forall t\ge 0, \quad \Exp [V_o(\rho_t)]\le \exp(-rt)V_o(\rho_0) $$ with rate $r=\tfrac{\eta}{2}\;\min_{k,k'}(\lambda_k-\lambda_{k'})^2.$\newline
In this sense, the open-loop system \eqref{eq:QND_OL} converges, for all initial states, towards the set of invariant states described in point $(ii)$.				
			\end{enumerate}
\end{lem}
\vspace{2mm}

\noindent \textbf{Proof:} The fact that $\mathcal{S}$ is positively invariant for \eqref{eq:QND_OL} is standard~\cite{barchielli2009quantum}.  

(i) For each $k$, the subspace populations $\bP_{k}(\rho)$ follow the It\={o} SDE:
  \begin{equation*}
  d\bP_k(\rho_t)=2\sqrt{\eta}\left(\lambda_k-\sum_{k'=1}^d\lambda_{k'}\bP_{k'}(\rho_t)\right)\bP_k(\rho_t) dW_t\;.
  \end{equation*}
  Taking the expectation yields $\tfrac{d}{dt}\Exp[\bP_{k}(\rho_t)]=0$, so indeed $\Exp[\bP_{k}(\rho_t)]=\bP_{ k}(\rho_0)$, $\forall t \ge 0$.

(ii) Take $\rho_0$ such that $\bP_{k}(\rho_0)=1$. Plugging into~\eqref{eq:QND_OL} we have $\cD_{L}(\rho_0)=L\rho_0L-\tfrac{1}{2}L^2\rho_0-\tfrac{1}{2}\rho_0L^2=\lambda_k^2\rho-\tfrac{1}{2}(2\lambda_{k}\rho_0)=0$ and $\cM_{L}(\rho_0)=\sqrt{\eta}(L\rho_0+\rho_0L-\tr{2L\,\rho_0}\rho_0)=\sqrt{\eta}(2\lambda_k\rho_0-\tr{2\lambda_k\rho_0}\rho_0)=0$. Thus $\rho_0$ is a steady state of~\eqref{eq:QND_OL}.

(iii) $\;V_o$ is a positive definite function on $\mathcal{S}$ and it equals $0$ only when $\bP_{ \ell}(\rho)=1$ for some $\ell$. It remains to check that it is a supermartingale with exponential decay. By It\={o}'s formula \eqref{eq:ItoFormula}, the variable $\xi_k:=\sqrt{\bP_k}$ satisfies
  \begin{equation}\label{eq:dxiell}
  d\xi_{k}=-\tfrac{1}{2}\eta(\lambda_{k}-\varpi(\xi))^2 \xi_k dt +  \sqrt{\eta}(\lambda_{k}-\varpi(\xi))\xi_{k}dW,
  \end{equation}
  with $\varpi(\xi)=\sum_{k=1}^d \lambda_{k} \xi_{k}^2$.

 Since $V_o(\rho)= \sum_{1\leq k < k'\leq d}^d  \xi_{k}\xi_{k'}$, consider the computation with It\={o}'s formula:
  \begin{multline*}
  d(\xi_k\xi_{k'})=(d\xi_k)\xi_{k'}+\xi_k(d\xi_{k'})+(d\xi_k)(d\xi_{k'})\\
  =-\tfrac{1}{2}\eta(\lambda_{k}-\varpi(\xi))^2 \xi_k \xi_{k'}dt +  \sqrt{\eta}(\lambda_{k}-\varpi(\xi))\xi_{k}\xi_{k'}dW\\
  \phantom{==}-\tfrac{1}{2}\eta(\lambda_{k'}-\varpi(\xi))^2 \xi_k\xi_{k'} dt +  \sqrt{\eta}(\lambda_{k'}-\varpi(\xi))\xi_{k}\xi_{k'}dW\\	  +\eta(\lambda_{k}-\varpi(\xi))(\lambda_{k'}-\varpi(\xi))\xi_{k'}\xi_{k}dt\\
  =-\tfrac{1}{2}\eta(\lambda_k-\lambda_{k'})^2\xi_{k'}\xi_{k}dt+\sqrt{\eta}(\lambda_{k}+\lambda_{k'}-2\varpi(\xi))\xi_{k}\xi_{k'}dW.
  \end{multline*}
The Markov generator $\mathcal{A}$ (see~\eqref{App:DiffOp}) following from the expectation of this equation thus yields:
  $$
  \mathcal{A} V_o =- \tfrac{\eta}{2} \sum_{k'=1}^d \sum_{k'<k}   ( \lambda_{k}-\lambda_{k'} )^2  \;\xi_{k} \xi_{k'} \; .
  $$
Since each $\xi_k(t)$ remains non-negative for all $t$, this readily yields
  $$
  \mathcal{A} V_o  \leq -   \tfrac{\eta}{2} \left(\min_{k',k\neq k'} (   \lambda_{k}-\lambda_{k'})^2 \right) V_o.
  $$
  By Theorem \ref{thm:ExponentialDecay} in Appendix, $V_o$ decays exponentially towards zero, concluding the proof.
  \hfill $\square$\\

The open-loop system corresponding to a QND measurement (\ref{eq:QND_OL}),(\ref{eq:QND_Y}) thus converges towards an eigenstate of $L$, satisfying $\bP_k(\rho)=1$ for some $k$, for each realization. However, the particular eigenstate $k$ will be random, with correlated measurement results indicating which state has been chosen; from (i), the probability to converge towards the particular eigenstate $k$ is equal to $\bP_k(\rho_0)$. 

The control objective is to ensure convergence to a \textit{target QND eigenstate}, indexed by $\ell\in\{1,\ldots,d\}$, for all realizations. More precisely, we will design a continuous stochastic real feedback process $v$, depending on the state $\rho$, such that $\lim_{t\rightarrow\infty}\Exp[\bP_{ \ell}(\rho_t)]=1$ with exponential convergence rate, for any initial condition $\rho_0\in\mathcal{S}$. The feedback action is modeled by means of a unitary control operation $U_t=e^{-iHudt}$ during the infinitesimal interval $[t,t+dt]$, where $H=H^\dagger$ is the actuator Hamiltonian and $dv=u dt$ is the feedback input signal.  Applied  on \eqref{eq:QND_OL}, the \textit{closed-loop} system reads then:
\begin{equation}\label{eq:QND_CL}
  \rho_{t+dt}=e^{-iHudt}(\rho_t+d\rho_t) e^{iHudt}
 \end{equation}
with measurement process $Y$ still given by~\eqref{eq:QND_Y}. Regarding exponential convergence, we aim to provide a global Lyapunov function $V(\rho)$ such that in closed-loop, $V(\rho_t)$ is a supermartingale with exponential decay for all $t\ge 0$ and all $\rho_0\in \mathcal{S}$. Providing feedback controls that ensure decay of $V$ in this sense has remained so far an open issue.

\section{Existing feedback designs}\label{section:PreviousWork}

Measurement-based quantum feedback has been investigated thoroughly in the literature, considering a wide array of applications such as state preparation \cite{stockton2004deterministic,mirrahimi2007stabilizing}, state purification \cite{combes2006rapid,wiseman2006reconsidering} or continous-time quantum error correction \cite{ahn2002continuous,sarovar2004practical}. Nevertheless, to our knowledge, control design has been done mainly on two feedback architectures: static output feedback \cite{wiseman2002bayesian} (so-called \textit{Markovian feedback} in the physics literature) and \textit{Bayesian feedback} \cite{doherty2000quantum,doherty1999feedback,mirrahimi2007stabilizing} involving a full state estimate. Intermediate approaches have attracted much less attention and promise an interesting research area.

\subsection{Static output feedback}
Static output feedback corresponds to quantum feedback of the form $dv=udt=fdt+\sigma dY$, where $f$ and $\sigma$ are constant. From~\eqref{eq:QND_CL} the dynamics in closed-loop read (see e.g. \cite{wiseman1994quantum,wang2001feedback}):
\begin{multline*}
d\rho =- i \left(f[ H~, \rho]+
\sqrt{\eta}\sigma [H~, ~L\rho + \rho L]\right)  dt \\
+ \big(  \cD_{L}(\rho) +\sigma^2\cD_{H}(\rho) \big) dt + \ \left( \sqrt{\eta} \cM_{L} (\rho)- i \sigma[ H~,~\rho]  \right) dW.
\end{multline*}
The simplicity of the feedback scheme makes it attractive for experimental implementations, since it avoids any overhead associated to dynamical computations in the feedback loop; in particular it needs no quantum state observer. With proper tuning of the constants $f,\sigma$, Markovian feedback allows to exponentially stabilize a range of target states, with direct algebraic proofs \cite{wiseman2002bayesian,wang2001feedback,ticozzi2008quantum,ticozzi2009analysis,ticozzi2010stabilizing}.

Unfortunately, an invariance argument shows that precisely the QND eigenstates, which are of particular interest in quantum engineering, are not stabilizable using this static output feedback with $f,\sigma$ just constants (\cite{cardona2018exponential}). 
 
It seems that a more involved controller is needed in order to bias the stochastic evolution towards a prescribed QND measurement eigenstate. One solution is to turn to state feedback, assuming an underlying quantum state observer.

\subsection{State feedback}

The standard state feedback takes the form $\;dv=udt= f(\rho) dt\;$. Thus the control signal is a deterministic scalar function of the state $\rho$. The closed-loop model for this quantum state feedback just takes the form:
\begin{equation*}\label{eq:CtrlDet}
	d\rho=-if(\rho)[H,\rho]dt+\cD_{L}(\rho)dt+\sqrt{\eta}\cM_{L}(\rho)dW.
\end{equation*}
Stabilization of QND eigenstates under this controlled dynamics has been treated extensively in the literature  \cite{ahn2002continuous,van2005feedback,yamamoto2007feedback,mirrahimi2007stabilizing,tsumura2007global,tsumura2008global,HadisCDC2018}. These results have succeeded in proving asymptotic convergence, in proper probabilistic settings. There is however room for improvement on a few aspects.

A first aspect is the convergence speed. Most papers do not provide any convergence rate; as the strongest result so far to our knowledge, \cite{HadisCDC2018} provide an estimate of the Lyapunov exponent for a qubit, valid for the final approach of the target state after an unspecified final initial transient. By Lemma \ref{lemma:OpenLoop}, the open-loop system under QND measurements converges towards the set of its steady states at global exponential speed. The absence of a proven similar property for the selection of one
target QND steady state thus appears as an avoidable gap.

A second aspect is that full state feedback, as is used in the above proposals, may appear unnecessary for this application. Running a quantum state estimate in real-time does pose experimental challenges, given the very short timescales involved (nanoseconds) and the plan to ultimately scale quantum computers to high-dimensional systems. In the present task, the asymptotic behavior is directly visible on the measurement signal. Indeed, the measurement signal corresponding to $\bP_ {k}(\rho_t)=1$ for all $t\geq 0$ is $Y_t= Y_0+  2\lambda_k  t + W_t$, with thus an expectation that directly informs on the eigenstate via the drift $\lambda_k\, t$, and a standard deviation in $\sqrt{t}$. This suggests that simple filtering should allow to essentially solve the task too. The necessity of keeping a full state observer, including quantum coherences, comes from the actuation strategy. This could be improved.

A third aspect, more related to model uncertainties, is that many existing proposals work with short pulses, inspired from the discrete-time counterpart. In actual implementations, it may be more cautious to use smoother control signals in order to avoid exciting spurious dynamics.\\

With respect to these two main approaches, we thus aim for an intermediate solution using a reduced estimator and smoother controls, while providing an exponential convergence guarantee. For this we resort to a control signal with a novel structure, called noise-assisted feedback.

\section{Noise-assisted feedback scheme}\label{section:NoiseAssistedFeedback}

In our feedback scheme, the control signal $dv=u dt$ still depends on (part of) the state like in the Bayesian feedback approach, but instead of just involving a deterministic function we drive $dv$ by an exogenous Brownian noise $dB$ independent of $W_t$:
\begin{equation}\label{eq:noiseFB}
dv_t=u_tdt=\sigma(\rho_t)dB_t \; .
\end{equation}
The feedback control occurs by making the noise gain $\sigma(\rho)$ a continuously differentiable function of $\rho$. Using in \eqref{eq:QND_CL} the Baker-Campbell-Hausdorff (BCH) formula $
e^{xH}\rho e^{-xH}=\sum_{j}T_j \frac{x^j}{j!},
$  where  $T_0=\rho$ and $T_{j+1}=[H,T_{j}]$ for $j\geq 0$, 
and applying the It\={o} rules $(dv)^2=\sigma^2(\rho)dt$, $dtdv=dtdB=dWdB=dt^2=0$, yields the following closed loop dynamics with two independent Wiener processes $W$ and $B$ :
\begin{multline}\label{eq:ControlNoise}
d\rho= \left(\cD_{L}(\rho)+ \sigma(\rho)^2\cD_{H}(\rho)\right)  dt +\sqrt{\eta} \cM_{L}(\rho)~ dW + \sigma(\rho)  i[\rho,H] ~dB.
\end{multline}

The main idea comes down to noise discouraging the system to converge towards an eigenstate different from $\bP_ {\ell}(\rho)=1$. Indeed, as the open-loop QND dynamics stochastically converges to one of the QND eigenstates, but on the average does not move closer to any particular one, it is sufficient to activate noise only when the state is close to a bad equilibrium in order to ``shake it away'' and induce global convergence to the target. Accordingly, we consider the feedback~\eqref{eq:noiseFB} with the following gain law:
\begin{equation}\label{eq:sigma}
  \sigma(\rho)= \bar \sigma ~ \varphi\left(\frac{\max_{k\neq\ell}\bP_k(\rho)-\pmin}{\pmax-\pmin}\right)
\end{equation}
where $\varphi\geq 0$ is a smooth saturating function on $[0,1]$, i.e. $\varphi(]-\infty,0])=\{0\}$ and $\varphi([1,+\infty[)=\{1\}$, with parameters $\bar\sigma >0$ and
$1 > \pmax > \pmin > \frac{1}{2}$. Since $\sum_{k} \bP_k=1$, each $\bP_k\geq 0$ and $\pmin>1/2$, the argument of the max can only change when $\max_{k\neq\ell}\bP_k(\rho)-\pmin < 0$. Therefore, the function $\rho\mapsto \sigma(\rho)$ is smooth despite the use of a max in its definition.

\section{Exponential stabilization via noise-assisted feedback }\label{section:ExpoStab}

We now construct a Lyapunov function and prove that our feedback design ensures its exponential convergence. A particular point for $n$-level systems, compared to $2$-level systems like in \cite{cardona2018exponential,HadisCDC2018}, is to take the limited actuation into account.

Inspired by~\cite{amini2011design}, we consider the $d\times d$ real symmetric matrix $\Delta$ with components
\begin{equation}\label{eq:LaplacianMatrix}
  \Delta_{k,k'}= \tr{\Pi_k \mathcal{D}_H(\Pi_{k'})} \; ,
\end{equation}
combining  the spectral decomposition $L=\sum_k \lambda_k \Pi_k$ with the actuator Hamiltonian $H$.  Its off-diagonal elements are non negative  since
$ \Delta_{k,k'}= \tr{\Pi_k H \Pi_{k'} H} \geq 0$  for $k\neq k'$. Its diagonal elements are non positive and
\begin{multline*}
  \Delta_{k,k}= \tr{\Pi_k H \Pi_k H} - \tr{\Pi_k H^2}
  \\= \tr{\Pi_k H \Pi_k H} - \tr{\Pi_k H (\Pi_1+\ldots+\Pi_d) H}
  \\ = - \sum_{k'\neq k} \tr{\Pi_k H \Pi_{k'} H}
  .
\end{multline*}
Thus $\Delta$ is a Laplacian matrix.

\begin{thm}\label{thm:QNDfeedback}
Assume that $L$ is nondegenerate, i.e., $d=n$ and each projector $\Pi_k$ is a rank one projector.
Consider the closed-loop system~\eqref{eq:ControlNoise} with feedback gain $\sigma(\rho)$ given by~\eqref{eq:sigma} for a given projector  $\Pi_\ell$  with  $\ell\in\{1,\ldots, d\}$.  Assume that the graph associated to the Laplacian matrix $\Delta$ defined in~\eqref{eq:LaplacianMatrix} is connected.  Then there exists $\overline{p} \in ] \tfrac{1 }{2}, 1[$ such that for any choice of  parameter $\bar\sigma >0$ and parameters  $1>\pmax > \pmin \geq \overline{p}$, the closed-loop trajectories converge exponentially to $\rho=\Pi_\ell$, in the sense that: there exist constants $\nu>0$ and $C >0$ (depending on $\bar\sigma$, $\pmax$, $\pmin$) for which 
$\;\Exp\left[\sqrt{1 - \bP_\ell(\rho_t)} \right] \leq C e^{-\nu t} \sqrt{1 - \bP_\ell(\rho_0)}$
for any initial state $\rho_0\in\mathcal{S}$.
\end{thm}
 If the  graph associated to $\Delta$ is not fully connected then there exists a partition of $\{1, \ldots, d\}=I \cup J$ ($I, J\neq \emptyset$, $I\cap J=\emptyset$) such that  $H=\Pi_I H \Pi_I + \Pi_J H \Pi_J$ with $\Pi_I=\sum_{k\in I} \Pi_k$ and $\Pi_J= \sum_{k\in J} \Pi_k$. Then any trajectory  $\rho_t$  of~\eqref{eq:QND_CL} with any feedback scheme  starting from $\tr{\rho_0 \Pi_I}=0$ (resp. $\tr{\rho_0 \Pi_J}=0$), satisfies $\tr{\rho_t \Pi_I}=0$ (resp. $\tr{\rho_t \Pi_J}=0$) for all $t >0$. Thus, closed-loop convergence to  $\bP_\ell=1$ with $\ell \in I$ is impossible when $\tr{\rho_0 \Pi_I}=\sum_{k\in I} \bP_k(\rho_0) < 1$.  In this sense the above connectivity condition on the graph of $\Delta$  cannot be weakened.

\vspace{2mm}

\noindent \textbf{Proof:} 
\newline
\emph{(-- Lyapunov function construction --)}
 We do not use directly $\sqrt{1-\bP_\ell(\rho)}$ as a closed-loop Lyapunov function. Instead we construct a closed-loop Lyapunov function $V(\rho)$ equivalent to $\sqrt{1-\bP_\ell}$ (i.e. $ c_* V(\rho) \leq \sqrt{1-\bP_\ell(\rho)} \leq  c^*  V(\rho) $ with $0 < c_*< c^*$)  such that $\mathcal{A} V \leq - r V$. More precisely, we use
 $$
 V_\alpha(\rho) = \sum_{s\in\{1,\ldots,d\}\setminus\{\ell\}} \sqrt{\sum_{k\in\{1,\ldots,d\}\setminus\{\ell\}} \alpha_{s,k}\; \bP_k(\rho) } \; .
 $$
The positive parameters $\alpha_{s,k}$ will be given by solving $d-1$ linear systems, indexed by $s$:
 $$
 \sum_{k'}\Delta_{k,k'} \alpha_{s,k'} = -\beta_{s,k}
 $$
 with $\beta_{s,k} >0$ for $k\neq \ell$ and $\beta_{s,\ell}= - \sum_{k\neq \ell} \beta_{s,k}$. Standard  arguments used  in~\cite{amini2011design}  guarantee under the connectivity assumption  that there exists, for each $s$, a unique solution $(\alpha_{s,k})$ such  that $\alpha_{s,k}>0$ for $k\neq \ell$ and $\alpha_{s,\ell}=0.$ (see, e.g.~\cite[Chapter 4]{beineke2004topicsgraphtheory}). When the $(d-1)\times d$ matrix $\beta$ is chosen of maximal rank $d-1$, the obtained matrix $\alpha$ is also of maximal rank $d-1$.
 Since $\sum_{k\in\{1, \ldots,d\}\setminus\{\ell\}}\bP_k=1-\bP_\ell$, one has
 $$
 c_* V_\alpha(\rho) \leq \sqrt{1-\bP_\ell(\rho)} \leq  c^*  V_\alpha(\rho)
 $$
 where  $c_*=\frac{1}{(d-1)\sqrt{\alpha^* }}$ and $c^*=\frac{1}{ (d-1)\sqrt{\alpha_*}}$ with
 $$
 \alpha_*=\min_{s,k\in\{1, \ldots,d\}\setminus\{\ell\}}\alpha_{s,k} \text{ and } \alpha^*=\max_{s,k\in\{1, \ldots,d\}\setminus\{\ell\}}\alpha_{s,k}
 .
 $$

\noindent \emph{(-- expression of the stochastic Markov generator --)}
The rest of the proof consists in showing that for any such choice of maximal rank matrix $\beta$, the resulting $V_\alpha$ becomes an exponential Lyapunov function as soon as  $\overline p$ is close enough to $1$ and $\pmin > \overline{p}$. This is based on the following  simple but slightly tedious computations of $\mathcal{A}V_\alpha$:
\begin{equation}\label{eq:AV}
  \mathcal{AV_\alpha}(\rho) = \tfrac{\sigma^2(\rho)}{2}  f_\alpha(\rho) - \tfrac{\eta}{2} g_\alpha(\rho) - \tfrac{\sigma^2(\rho)}{8} h_\alpha(\rho)
\end{equation}
with
\begin{align*}
  f_\alpha(\rho)&= \sum_{s} \frac{\Sigma_k \alpha_{s,k}\tr{\Pi_k \mathcal{D}_{H}(\rho)} }{\sqrt{\Sigma_k \alpha_{s,k} \bP_k}}
  \\
  g_\alpha(\rho)&=  \sum_{s} \left( \frac{\Sigma_k \alpha_{s,k} (\lambda_k - \tr{L\rho}) \bP_k}{\Sigma_k \alpha_{s,k} \bP_k}\right)^2  \sqrt{\Sigma_k \alpha_{s,k} \bP_k}
   \\
   h_\alpha(\rho)&= \sum_{s} \frac{\left(\Sigma_k \alpha_{s,k} \tr{i[\Pi_k,H]\rho}\right)^2}{\left(\Sigma_k \alpha_{s,k} \bP_k\right)^{3/2}}
   .
\end{align*}
These expressions are obtained with the following general formula based on It\={o} rules ($a_k>0$ constant)
$$
  d \sqrt{\sum_k a_k \bP_k} = \frac{\sum_k a_k d \bP_k}{2\sqrt{\sum_k a_k \bP_k}}
  -  \frac{\left(\sum_k a_k d \bP_k\right)^2}{8\left(\sum_k a_k \bP_k\right)^{3/2}}
$$
 with $
 \Exp[d\bP_k | \rho] = \sigma^2(\rho) \tr{\Pi_k \mathcal{D}_{H}(\rho)} ~dt
 $
 and
$$
   \Exp[(\Sigma_k a_k d\bP_k)^2 | \rho]
   =
     \sigma^2 \left( \Sigma_k a_k \tr{i[\Pi_k,H]\rho} \right)^2 dt     
      +4 \eta \left( \Sigma_k a_k (\lambda_k - \tr{L \rho}) \bP_k \right)^2 dt
      .
$$
 
 \noindent \emph{(-- closed-loop essential contribution --)} For $p\in[0,1]$ let
 $$
 \mathcal{S}_{p} \triangleq \left\{ \rho\in\mathcal{S}~|~\exists j \neq \ell,~\bP_j(\rho) \geq p \right\}
 .
 $$
 Take $\rho \in \mathcal{S}_p$ with $p=1$. Then there exists  $j\in\{1,\ldots,d\}\setminus\{\ell\}$ such that  $\rho=\Pi_j$, thus
 $\sum_k \alpha_{s,k} \bP_k(\rho)=\alpha_{s,j} >0$ and 
 $$
 \Sigma_k \alpha_{s,k}\tr{\Pi_k \mathcal{D}_{H}(\rho)} = -\beta_{s,j} < 0.
 $$
 Consequently $f_\alpha(\rho)= -\sum_{s\in\{1,\ldots,d\}\setminus\{\ell\}}\frac{\beta_{s,j}}{\sqrt{\alpha_{s,j}}} < 0$ with $V_\alpha(\rho)=\sum_{s} \sqrt{\alpha_{s,j}} >0$.
 
 By continuity of $f_\alpha/V_{\alpha} $ on $\mathcal{S}_{1/2}$, there exist $\epsilon_f >0$ and $\overline{p}\in]1/2,1[$ such that
 $$
 f_\alpha(\rho) \leq - \epsilon_f V_\alpha(\rho) \quad \text{for all  } \rho \in \mathcal{S}_{\overline{p}}
 .
 $$
Taking $\pmin > \overline{p}$, we ensure that the feedback will only be turned on when it contributes a negative term to $\mathcal{AV}_\alpha$.\vspace{3mm}
 
 \noindent \emph{(-- open-loop essential contribution --)} For all $\rho\in\mathcal{S}\setminus\{\Pi_\ell\}$,  $\chi(\rho)= g_{\alpha}(\rho)/V_\alpha(\rho)$ is well defined.  Since $\tr{\rho L}=\sum_{k} \lambda_k \bP_k(\rho)$, the function $\chi(\rho)$ depends only on the populations $\bP_k$. Consider the following parametrization exploiting the degree 0 homogeneity of $\chi$ in the populations:
 $$
 r=1-\bP_\ell,\quad x_k = \bP_k / (1-\bP_\ell) \text{ for } k\neq \ell
 .
 $$
 For $\rho\in\mathcal{S}\setminus\{\Pi_\ell\}$, the function  $\chi$ admits the following smooth expression with the variables $r\in]0,1]$ and $x_k\in[0,1]$ satisfying $\sum_{k\neq \ell} x_k=1$:
 $$
 \chi(r,x)=
 \frac{
  \sum_{s\neq \ell} \left( \tfrac{\Sigma_{k\neq\ell} \alpha_{s,k} (\lambda_k - \varpi(r,x) )x_k}{\Sigma_{k\neq\ell} \alpha_{s,k} x_k}\right)^2  \sqrt{\Sigma_{k\neq\ell} \alpha_{s,k} x_k}
  }
  {
  \sum_{s\neq\ell}  \sqrt{\Sigma_{k\neq\ell} \alpha_{s,k} x_k}
  }
  $$
  with $\varpi(r,x)=  (1-r)\lambda_\ell + r \left(\sum_{k\neq\ell} \lambda_k x_k\right) $. We study its extension to the compact set with $r\in[0,1]$. Clearly $\chi(r,x)\geq 0$.   Consider the solutions  $(r,x)$  of $\chi(r,x)=0$. Necessarily, they satisfy
  $$
  \forall s\neq \ell, \quad \sum_{k\neq\ell} \alpha_{s,k} (\lambda_k - \varpi(r,x) )x_k=0
  .
  $$
  Since the $(d-1)\times d$ matrix $\alpha_{s,k}$ is  of maximal rank $d-1$,
  $ \forall k\neq \ell $ we have  $(\lambda_k - \varpi(r,x) )x_k=0.$
  Taking the sum over $k\neq\ell$, one gets
  $
  (1-r)\left(\lambda_\ell - \sum_{k\neq\ell} \lambda_k x_k\right) =0
  $
  implying two possibilities:
  \begin{itemize}
    \item if $\lambda_\ell = \sum_{k\neq\ell} \lambda_k x_k$, then $\varpi(r,x)=\lambda_\ell$ and the condition before summing requires $(\lambda_k - \lambda_\ell )x_k=0$ for $k\neq \ell$. Since $\lambda_k\neq \lambda_\ell$ this implies that $x_k=0$ $\forall k\neq \ell$. But this is not possible since $\sum_{k\neq \ell}x_k=1$.
    \item if $r=1$, then $\varpi(r,x)=\sum_{k'\neq \ell} \lambda_{k'} x_{k'}$. For $k\neq \ell$ the conditions become $\left(\lambda_k - \sum_{k'\neq \ell} \lambda_{k'} x_{k'}\right)x_k=0$ before summing. Since $\lambda_k\neq \lambda_{k'}$ for $k\neq  k'$ and $x_k\in[0,1]$ with $\sum_{k\neq \ell}x_k=1$, there necessarily exists $j\neq \ell$ such that $x_j=1$ and $x_k=0$ for $k\notin\{\ell,j\}$.
  \end{itemize}
  Consequently, the nonnegative smooth function $\chi(r,x)$ vanishes only at $d-1$ isolated points, where $r=1$ and $x_k=\delta_{kj}$ for some $j\neq \ell$.  By continuity, for any $p \in ]0,1[$, there exists  $\theta_p>0$ such that
  $  \forall \rho\in\mathcal{S}\setminus(\mathcal{S}_p \cup \{\Pi_\ell\})$, we have $\chi(\rho) \geq \theta_p$.  This proves that
  $$
  g_\alpha (\rho) \geq  \theta_p V_\alpha(\rho) \quad \forall \rho\in\mathcal{S}\setminus\mathcal{S}_p\; .
  $$
  
\noindent \emph{(-- bringing all pieces together --)} To conclude, consider $\mathcal{A} V_\alpha$ given in~\eqref{eq:AV}. Since $h_\alpha \geq 0$, we have
   $$
  \forall \rho \in \mathcal{S}, \quad  \mathcal{AV_\alpha}(\rho) \leq \tfrac{\sigma^2(\rho)}{2}  f_\alpha(\rho) - \tfrac{\eta}{2} g_\alpha(\rho) .
   $$
Consider the feedback gain $\sigma(\rho)$  with $\pmin > \overline{p}$. Then from the closed-loop essential contribution,
$$
\mathcal{AV_\alpha}(\rho) \leq - \tfrac{\bar\sigma^2}{2} \epsilon_f V_\alpha(\rho) \quad \forall \rho \in \mathcal{S}_{\pmax}
$$
and from the open-loop essential contribution,
$$
\mathcal{AV_\alpha}(\rho) \leq -\tfrac{\eta}{2}  \theta_{\pmax}  V_\alpha(\rho) \quad \forall \rho \in \mathcal{S}/ \mathcal{S}_{\pmax} \; .
$$
Thus for all
$ \rho$ in $ \mathcal{S}$, one has  $\mathcal{AV_\alpha}(\rho) \leq -\nu  V_\alpha(\rho)$ where $\nu$ is equal to $\min\left(  \tfrac{\bar\sigma^2\epsilon_f}{2} , \tfrac{\eta \theta_{\pmax}}{2} \right) $. A direct application of  Theorem~\ref{thm:ExponentialDecay} recalled  in Appendix ensures
 $\Exp(V_\alpha(\rho_t)]\leq  V_\alpha(\rho_0) e^{-\nu t}$.  
\hfill $\square$\\

\section{Observer and approximate quantum filtering} \label{section:Qfilters}

The feedback is based on the value $\rho_t$ of the quantum state, which is not directly measured. One has  to reconstruct in real-time this quantum state via a quantum filter, a quantum state-observer. For~\eqref{eq:ControlNoise}, it reads~(see e.g.~\cite{bouten2007introduction}):
\begin{multline*}
d\rho_t= \cD_{L}(\rho_t) dt +\sqrt{\eta} \cM_{L}(\rho_t) \big(dY_t-2\sqrt{\eta}\tr{\rho_t L }dt\big)\\ + \sigma(\rho_t)^2\cD_{H}(\rho_t)dt - i\sigma(\rho_t)[H, \rho_t]dB_t.
\end{multline*}
A standard result~\cite{stockton2004deterministic,van2005feedback,mirrahimi2007stabilizing,AminiSDSMR2013A}  ensures that  the resulting observer/controller closed-loop system
\begin{align*}
d\rho_t& = \cD_{L}(\rho_t) dt +\sqrt{\eta} \cM_{L}(\rho_t) dW_t
 \\ & \qquad + \sigma(\hrho_t)^2\cD_{H}(\rho_t)dt - i\sigma(\hrho_t)[H, \rho_t]dB_t.
 \\
 dY_t &= 2\sqrt{\eta}\tr{\rho_t L }dt + dW_t
 \\[0.5em]
d\hrho_t& = \cD_{L}(\hrho_t) dt +\sqrt{\eta} \cM_{L}(\hrho_t) \big(dY_t-2\sqrt{\eta}\tr{\hrho_t L }dt\big)
 \\ & \qquad + \sigma(\hrho_t)^2\cD_{H}(\hrho_t)dt - i\sigma(\hrho_t)[H, \hrho_t]dB_t.
\end{align*}
converges almost surely towards the stationary state $\rho=\hrho=\Pi_\ell$ as soon as $\tr{\rho_0 \hrho_0} >0$ and the convergence conditions of Theorem~\ref{thm:QNDfeedback} are satisfied.

Practical implementation of such quantum filter could be a problematic issue: when the dimension $n$ of the system is large, it requires  to store and update in real-time the $n(n-1)/2$ components of the operator $\hrho$.  Since the feedback depends only on the populations, i.e., the diagonal of $\rho$ in the eigenbasis of the measurement operator $L$, the development of a reduced-order (possibly approximate) quantum filter depending only on these populations is suggested.

A  first reduction consists in replacing  $\rho_t$ in the feedback law by $\widehat{\varrho}_t$ corresponding to the Bayesian estimate of $\rho_t$ knowing $\rho_0$ and $Y_{\tau}$ for $\tau\in [0,t]$.  One thus discards here the knowledge of  $B_t$. Then, one can prove that  $\widehat{\varrho}_t$ obeys to the following  stochastic differential  equation:
\begin{equation}\label{eq:CtrlFiltRed}
d\widehat{\varrho}_t= \cD_{L}(\widehat{\varrho}_t) dt +\sqrt{\eta} \cM_{L}(\widehat{\varrho}_t) \big(dY_t-2\sqrt{\eta}\tr{\widehat{\varrho}_t L }dt\big) + \sigma(\widehat{\varrho}_t)^2\cD_{H}(\widehat{\varrho}_t)dt
.
\end{equation}
This filter involves less computations but still the full matrix $\widehat{\varrho}$. Since the feedback law only depends on the $\bP_k(\widehat{\varrho})=\tr{\widehat{\varrho}\Pi_{k}}$, it would be ideal to have a reduced filter involving only those variables. In cases like \cite{cardona2019QEC}, the filter \eqref{eq:CtrlFiltRed} can be directly and exactly reduced to an autonomous system on the $\bP_k(\widehat{\varrho})$, as we discard the \emph{random coherences} among eigenstates induced by $dB_t$. However, when a single noise process $dB_t$ drives many levels at once, the filter \eqref{eq:CtrlFiltRed} has to model \emph{correlations among the various random coherences} and this precludes an exact reduction. Nevertheless, by neglecting those feedback-induced correlations (recalling that feedback actuation is often turned off), we can set coherences $\bket{k|\widehat{\varrho}|k'}$, $k\neq k'$ to zero, and propose an approximate filter for populations $\widehat{\bP}_k$ to estimate $\bP_k(\widehat{\varrho})$. This amounts to replacing $\cD_{H}(\widehat{\varrho})$ by a population transfer via the Laplacian matrix $\Delta$ defined in~\eqref{eq:LaplacianMatrix}. It yields:
\begin{equation}\label{eq:ReducedFilterGeneral}
d\widehat{\bP}_k=
2\sqrt{\eta} \widehat{\bP}_k \big(\lambda_k-\varpi(\widehat{\bP}) \big)
 \big(dY_t-2\sqrt{\eta}\varpi(\widehat{\bP}) dt \big)
 +\sigma^2(\widehat{\bP}) \sum_{k'=1}^{d}\Delta_{k,k'}\, \widehat{\bP}_{k'} \,dt
\end{equation}
where $\varpi(\widehat{\bP})=\sum_{k'=1}^{d}\lambda_{k'}\widehat\bP_{k'}$. This approximate filter requires to store and update in real-time only $d$ real numbers. For any measurement trajectory $Y_t$, the components of $\widehat{\bP}$ remain nonnegative and their sum equal to one. In open-loop ($\sigma\equiv 0$), this population filter is exact.

\section{Simulation and robustness  issues}\label{section:Simulations}
Theorem~\ref{thm:QNDfeedback} ensures for $\pmin$ close enough to $1$ exponential closed-loop convergence of~\eqref{eq:QND_CL} with noise-assisted feedback~(\ref{eq:noiseFB}),(\ref{eq:sigma}). This section is devoted to numerical estimation of closed-loop convergence rates and investigation of the related robustness on a specific quantum system already considered in~\cite{mirrahimi2007stabilizing,stockton2004deterministic}: a spin $J$ system of dimension $n=2J+1$ where the measurement operator is
$$
L=\sum_{m=0}^{2J} (J-m) \ket{J\minou m}\bra{J\minou m}
$$
and the actuator Hamiltonian is a tridiagonal matrix 
$$
H= \sum_{m=0}^{2J-1} \tfrac{\sqrt{(m+1)(2J-m)}}{2 i} \left(\ket{J\minou m}\bra{J\minou m\minou 1} - \ket{J\minou m\minou 1}\bra{J\minou m}\right)
$$
The Hilbert space is spanned by the $2J+1$ orthonormal vectors  $\ket{J\minou m}$  for $m=0, \ldots, 2J$.

All the simulations below correspond to detection efficiency $\eta=0.8$ and $J=2$ ($n=5$), for which: 
$$
L=\begin{pmatrix}
    2 & 0 & 0 & 0 & 0 \\
    0 & 1 & 0 & 0 & 0 \\
    0 & 0 & 0 & 0 & 0 \\
    0 & 0 & 0 & \minou 1 & 0 \\
    0& 0 & 0 & 0 & \minou 2 \\
  \end{pmatrix}
  , \quad 
  H=\frac{1}{2 i}\begin{pmatrix}
      0 & \minou2 & 0 & 0 & 0 \\
      2 & 0 & \minou\sqrt{6} & 0 & 0 \\
      0 & \sqrt{6} & 0 & \minou\sqrt{6} &0\\
      0 & 0 & \sqrt{6} & 0 & \minou2 \\
      0 & 0 & 0 & 2 & 0 \\
    \end{pmatrix}
    . 
$$
According to Lemma~\ref{lemma:OpenLoop}, the open-loop convergence rate is $\eta/2=0.4$. As a generic control goal we choose to stabilize the state $\Pi_\ell=\ket{0}\bra{0}$ associated to the zero eigenvalue of $L$.   For the noise-assisted feedback gain~\eqref{eq:sigma}, we take
$\bar \sigma= \sqrt{5\eta}$, $\pmax= \pmin+0.05$,  $\pmin=0.9$ or $\pmin=0.6$ with the saturation function $\varphi(s)=\min(1,\max(0,s))$.  For each case we simulate a set of 1000 realizations starting from the fully depolarized state $\rho_0= \bI/5$. We estimate the evolution of $\Exp[\sqrt{1-\tr{\Pi_\ell \rho_t}}]\equiv \Exp[\sqrt{1-\bket{0|\rho_t|0}}]$ by taking the ensemble average over these 1000 realizations. The Laplacian matrix
$$
\Delta = \begin{pmatrix}
           \minou 1 & 1& 0 & 0 & 0 \\
           1 & \minou \tfrac{5}{2} & \tfrac{3}{2} & 0 & 0 \\
           0 & \tfrac{3}{2} & \minou 3& \tfrac{3}{2} & 0 \\
           0 & 0 & \tfrac{3}{2} & \minou\tfrac{5}{2} & 1 \\
           0 & 0 & 0 & 1 & \minou 1 \\
         \end{pmatrix}
         ,
$$
inherits the tridiagonal structure of $H$ and thus admits a connected graph, so Theorem~\ref{thm:QNDfeedback} predicts exponential convergence with our controller using a sufficiently high value of $\pmin$ and the filter \eqref{eq:CtrlFiltRed}.

\begin{figure}
	\centering \includegraphics[width=100mm]{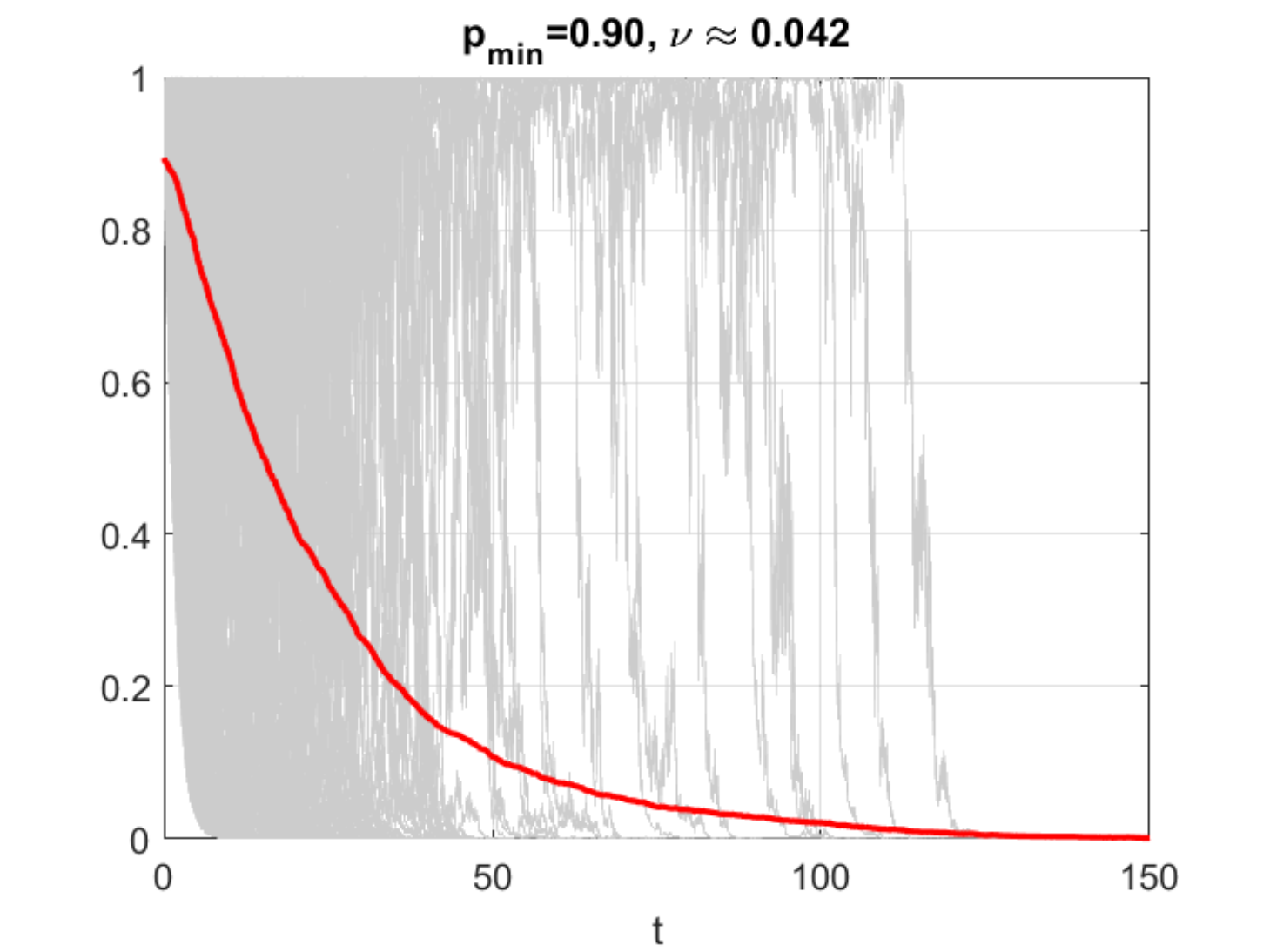}
	\caption{Ideal closed-loop simulations with $\pmin=0.9$.  In gray: selection of 200 individual trajectories $t\mapsto \sqrt{1-\tr{\Pi_\ell \rho_t}}$; In red: average over 1000 realizations, showing exponential convergence at a rate $\nu \approx 0.04$.}\label{fig:ideal09}
\end{figure}

\begin{figure}
	\centering \includegraphics[width=100mm]{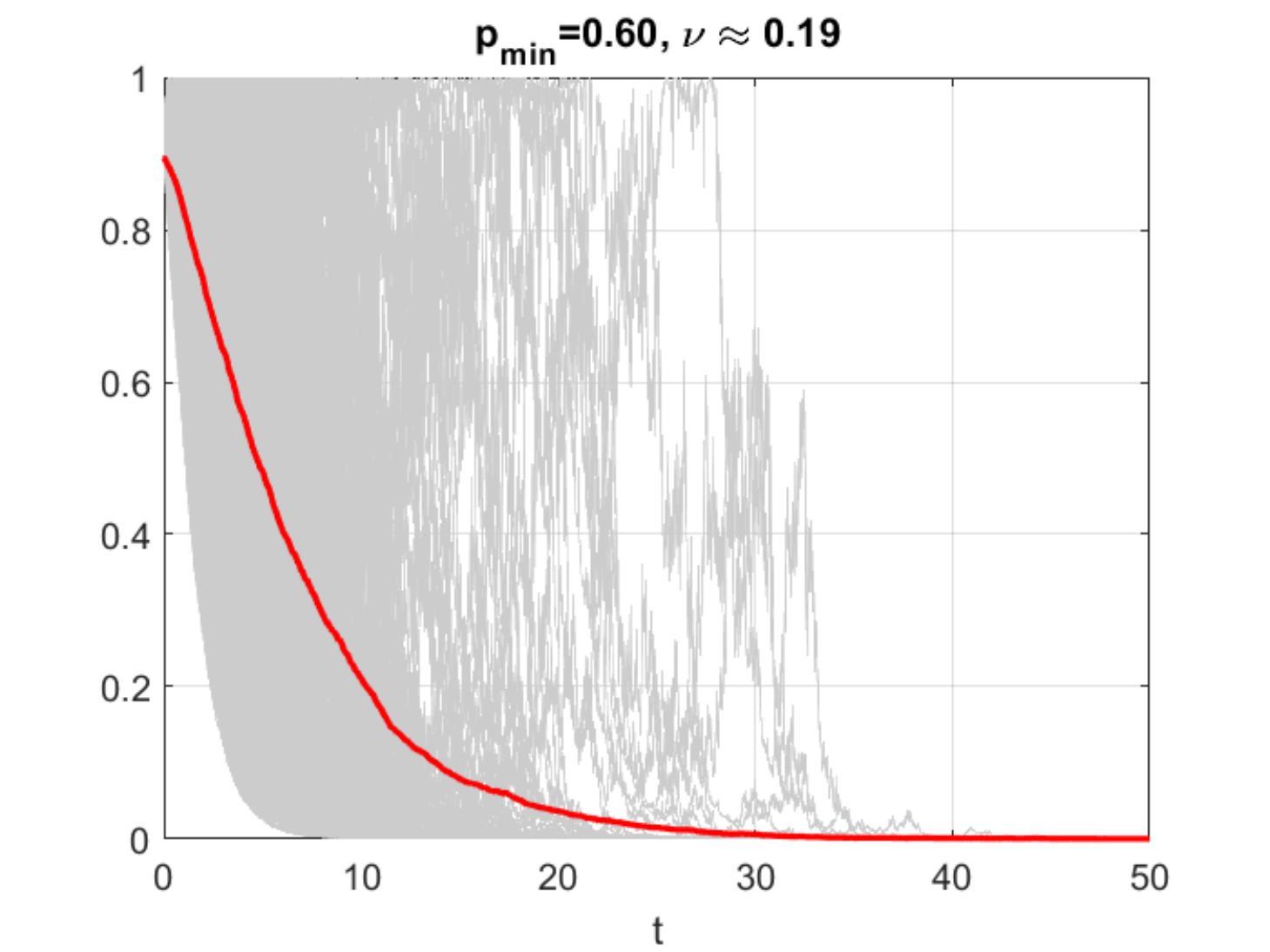}
	\caption{Ideal closed-loop simulations with $\pmin=0.6$.  In gray: selection of 200 individual trajectories $t\mapsto \sqrt{1-\tr{\Pi_\ell \rho_t}}$; In red: average over 1000 realizations, showing exponential convergence at a rate $\nu \approx 0.2$.}\label{fig:ideal06}
\end{figure}

We first illustrate the exponential convergence rate $\nu$. The proof of Theorem~\ref{thm:QNDfeedback} only provides a very loose bound on both $\nu$ and the necessary $\pmin$, so we here stick to numerical simulations. From the system analysis, the qualitative trend should be that lower values of $\pmin$ imply more frequent feedback corrections and thus faster convergence. Figure~\ref{fig:ideal09} illustrates a simulation set with $\pmin=0.9$, which fits an exponential convergence at rate $\nu\approx 0.04$. In a second simulation set, see Figure~\ref{fig:ideal06}, we have pushed this to $\pmin=0.6$ and observed a much faster convergence rate around $\nu \approx 0.2$, i.e.~one half of the open-loop convergence rate. This suggests that such noise-assisted feedback can be tuned to achieve convergence rates similar to the open-loop one.

\begin{figure}
	\centering \includegraphics[width=100mm]{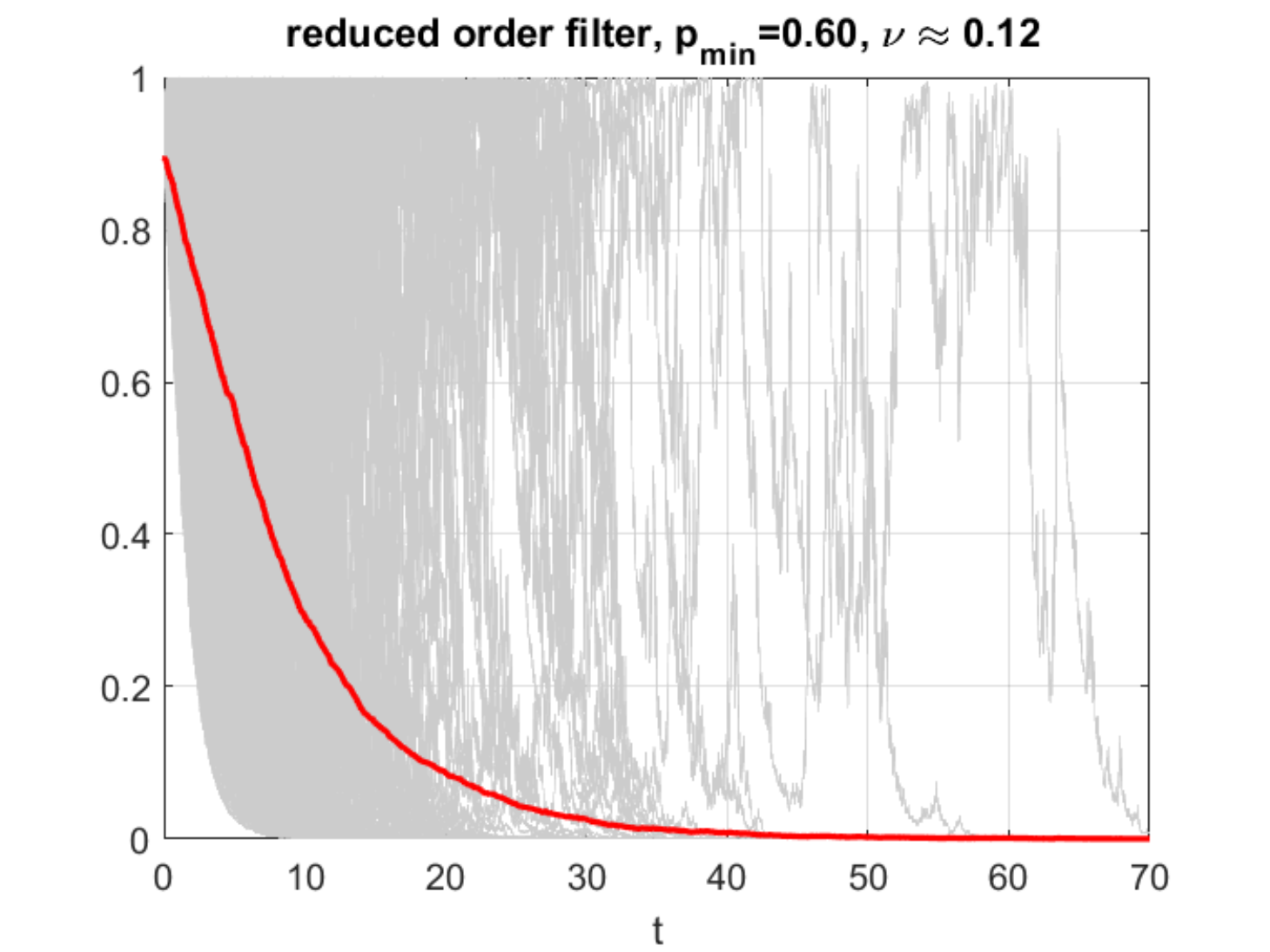}
	\caption{Closed-loop simulations with $\pmin=0.6$ with the approximate population filter~\eqref{eq:ReducedFilterGeneral}.  In gray: selection of 200 individual trajectories $t\mapsto \sqrt{1-\tr{\Pi_\ell \rho_t}}$; In red: average over 1000 realizations, showing exponential convergence at a rate $\nu \approx 0.12$.}\label{fig:reduced06}
\end{figure}

We next investigate the behavior with the approximate reduced filter on the population vector $\widehat{\bP}$. The simulations of Fig.~\ref{fig:reduced06} differ from the ones of Fig.~\ref{fig:ideal06} just by replacing in the feedback law the ideal populations $\bP$ by the approximated ones $\widehat{\bP}$ solutions of~\eqref{eq:ReducedFilterGeneral}. One observes still an exponential convergence, with a reasonable decrease of the convergence rate from $0.19$ to $0.12$, illustrating the practical interest of such low-dimensional filters. 

\begin{figure}
	\centering \includegraphics[width=100mm]{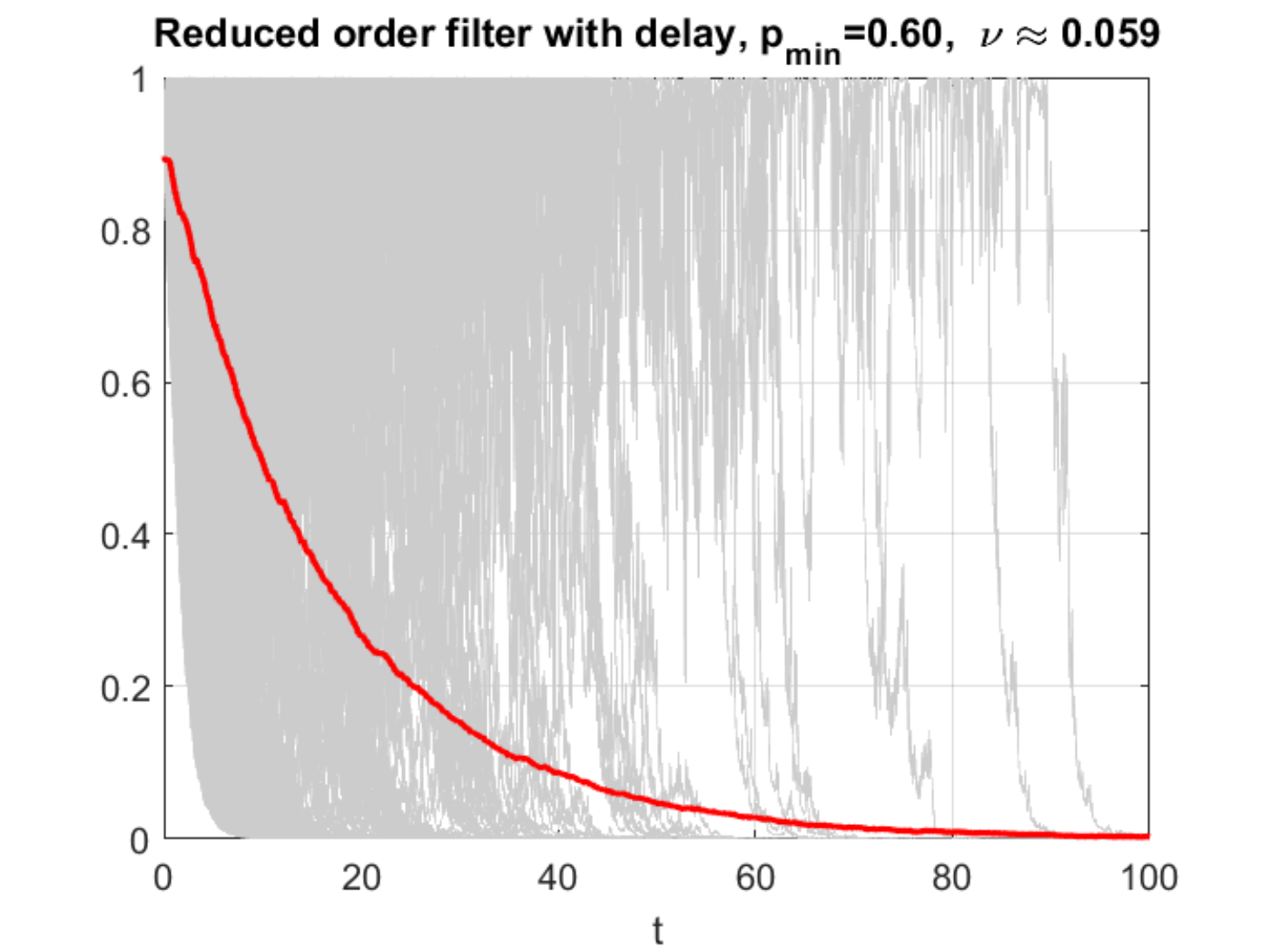}
	\caption{Closed-loop simulations with $\pmin=0.6$ with the approximate population filter~\eqref{eq:ReducedFilterGeneral} and feedback delay of $0.5$.  In gray: selection of 200 individual trajectories $t\mapsto \sqrt{1-\tr{\Pi_\ell \rho_t}}$; In red: average over 1000 realizations, showing exponential convergence at a rate $\nu \approx 0.06$.}\label{fig:delay06}
\end{figure}

Finally, on Fig.~\ref{fig:delay06} we further investigate robustness of the control strategy by using the reduced filter \eqref{eq:ReducedFilterGeneral} with a feedback delay of $0.5$~time units in the closed-loop simulations. Despite a decrease by a factor two of the convergence rate, the fact a feedback latency of $1/4$ of the open-loop convergence time does not destabilize this feedback scheme suggests promising robustness properties. 

\section{Concluding remarks}


We have approached the problem of stabilizing a QND measurement eigenstate in continuous-time by introducing Brownian noise to drive the control. The use of noise provides a simple controller that shakes away spurious steady states in closed loop and thus achieves exponential stabilization of a target eigenstate.

The present work still leaves room for improvement in, at least, the following directions: 
\begin{itemize}
\item While our proof of exponential convergence can provide an estimate of the convergence rate, we did not try to optimize the Lyapunov function nor the other control parameters in order to maximize the speed of convergence. We have shown numerically that the closed-loop convergence rate apparently can be made similar to the open-loop convergence rate with our approach, suggesting promising results in a precise analysis of convergence rate.
\item In numerical simulations, the reduced approximate filter \eqref{eq:ReducedFilterGeneral} appears good enough to achieve global exponential stabilization. We conjecture that this can be proven, for this filter and possibly for even simpler filters based on direct output signal filtering or sparse $\widehat{\bP}$.

\item The capability of performing several quantum measurements and of applying different unitary feedback controls on a single quantum system motivates the study of multi-input multi-output (MIMO) quantum feedback schemes. A MIMO version of static output feedback was introduced in \cite{chia2011quantum} and an implementation was made in  \cite{campagne2016using}. In addition, we have presented, in the context of quantum error correction \cite{cardona2019QEC}, a MIMO scheme of this noise-assisted feedback to stabilize a manifold of quantum states. A general MIMO extension of Theorem \ref{thm:QNDfeedback} should be feasible along the same lines as the present work. 
\end{itemize} 

\bibliography{root}
\bibliographystyle{plain}

\appendix

\section{Tools from stochastic stability}\label{section:StochasticStability}
We refer the reader to \cite{khasminskii2011stochastic} for further reference on these fundamental results. We consider concrete instances of It\={o} stochastic differential equations (SDEs) on $\mathbb{R}^n$ of the form
\begin{equation}\label{eq:SDE}
dx_t=\mu(x_t)dt+\theta(x_t)dW_t\; ,
\end{equation}
where $W_t$ is a standard Brownian motion on $\mathbb{R}^k$, and $\mu,\theta$ are regular functions of $x$ with image in $\mathbb{R}^n$ and $\mathbb{R}^{n\times k}$ respectively, satisfying the usual conditions for existence and uniqueness of solutions \cite{khasminskii2011stochastic} on $\mathcal{S}$, a compact and positively invariant subset of $\mathbb{R}^n$.

Let $\mathcal{I}:=\{x\in\mathcal{S}: \mu(x)=\theta(x)=0\}$ be an invariant set of \eqref{eq:SDE}. Let $V(x)$, a nonnegative real-valued twice continuously differentiable function on $ \mathcal{S}\setminus\mathcal{I}$, with $V(x)=0$ implying $x\in \mathcal{I}$. It\={o}'s formula on $V$ yields  \cite{oksendal2013stochastic}
\begin{equation}\label{eq:ItoFormula}
dV=\Big(\sum_i \mu_i\frac{\partial}{\partial x_i}V+\frac{1}{2} \sum_{i,j} \theta_i\theta_j\frac{\partial^2}{\partial x_ix_j}V\Big)dt +\sum_i \theta_i\frac{\partial}{\partial x_i}VdW_i.
\end{equation}
The Markov generator $\mathcal{A}$ of the SDE \eqref{eq:SDE} is defined for any function $V$ in its domain by
\begin{equation}\label{App:DiffOp}
\mathcal{A}V=\sum_i \mu_i\frac{\partial}{\partial x_i}V+\frac{1}{2} \sum_{i,j} \theta_i\theta_j\frac{\partial^2}{\partial x_ix_j}V.
\end{equation}
 It is related to the SDE \eqref{eq:ItoFormula} by \cite[Chapter 7]{oksendal2013stochastic}
$$
\Exp[V(x(t))]=V(x(0))+\Exp\left [\int_0^t \mathcal{A}V(x(s))ds\right].
$$

The stochastic counterpart of Lyapunov's second method provides sufficient conditions for stochastic stability by analyzing the generator $\mathcal{A}V$, e.g.:

\begin{thm}[Khasminskii \cite{khasminskii2011stochastic}]\label{thm:ExponentialDecay}
\textit{If for some $r>0$ the Markov generator satisfies $\mathcal{A}V(x)\le -rV(x)$ $\forall x\in \mathcal{S}$, then $\Exp[V(x(t))] \leq V(x(0))\; \exp(-r\, t)$ $ \forall t\ge 0$, i.e., $V(x(t))$ is a supermartingale with exponential decay. Thus $\lim_{t\rightarrow\infty}\Exp[V(x(t))]=0$; since $V(x)=0$ implies $x\in \mathcal{I}$, this implies convergence towards $\mathcal{I}$ of all the solutions of Eq. \eqref{eq:SDE} starting in $\mathcal{S}$. }
\end{thm}

\end{document}